\documentstyle[12pt,a4]{article}
\input tcilatex
\begin{document}

\title{Error threshold in the evolution of diploid organisms}
\author{D. Alves* and J. F. Fontanari \\
Instituto de F\'{\i}sica de S\~ao Carlos \\
Universidade de S\~ao Paulo \\
Caixa Postal 369 \\
13560-970 S\~ao Carlos SP \\
Brazil}
\date{}
\maketitle

\begin{abstract}
The effects of error propagation in the reproduction of diploid organisms
are studied within the population genetics framework of the quasispecies
model. The dependence of the error threshold on the dominance parameter is
fully investigated. In particular, it is shown that dominance can protect
the wild-type alleles from the error catastrophe. The analysis is restricted
to a diploid analogue of the single-peaked fitness landscape.
\end{abstract}


\bigskip


\vspace{1.0cm}

*e-mail: quiron@ifqsc.sc.usp.br

\vfill 

\begin{center}
IFSC-USP -  PREPRINT
\end{center}

\newpage


The finding that the length of self-reproducing molecules that compete for a
finite supply of resources is limited by their replication accuracy is
probably the main outcome of Eigen's quasispecies model (Eigen 1971). This
phenomenon, termed error threshold, poses an interesting challenge to the
theories of the origin of life, since it prevents the emergence of huge
molecules which could carry the necessary information for building a complex
metabolism (Eigen and Schuster 1979, Kauffman 1993).

In the quasispecies model, a molecule is represented by a string of $\nu$
digits $\vec{s} = \left (s_1, s_2, \ldots, s_\nu \right )$, with the
variables $s_i$ allowed to take on $\kappa$ different values, each
representing a different type of monomer used to build the molecule. The
focus is on the time evolution of the concentrations $x_i$ of molecules of
type $i =1, 2, \ldots, \kappa^\nu $ which obey the following differential
equations (Eigen 1971) 
\begin{equation}  \label{ODE}
\frac{dx_i}{dt} = \sum_j W_{ij} x_j - \left [ D_i + \Phi \left ( t \right )
\right ] x_i \; ,
\end{equation}
where the constants $D_i$ stand for the death probability of molecules of
type $i$ and $\Phi (t)$ is a dilution flux that keeps the total
concentration constant. The feature that distinguishes this model from the
well-established models of population genetics (Hartl and Clark 1989) is the
replication matrix $W $ which takes into account the primary structure of
the molecules. More specifically, its elements are given by 
\begin{equation}
W_{ii} = A_i \, q^\nu
\end{equation}
and 
\begin{equation}
W_{ij} = \frac{A_i}{\left ( \kappa - 1 \right )^{d \left( i,j \right )}} \,
q^{\nu - d \left ( i,j \right )} \left ( 1 - q \right )^{d \left ( i,j
\right )} ~~~~i \neq j ,
\end{equation}
where $A_i$ is the replication rate of molecules of type $i$, $d \left ( i,j
\right )$ is the Hamming distance between molecules $i$ and $j$, and $q \in
[0,1] $ is the single-monomer replication accuracy, which is assumed to be
the same for all monomers.

For simple replication landscapes, the solutions of the $\kappa^\nu$ kinetic
equations (\ref{ODE}) have been thoroughly studied using perturbation theory
(Eigen {\em et al} 1989). More complex, spin-glass-like replication
landscapes can be analysed using the correspondence between those equations
and the equilibrium properties of a surface lattice system (Leuth\"{a}user
1986, Leuth\"{a}user 1987, Tarazona 1992, Franz {\em et al} 1993). It is
worth to mention that for the single-peaked replication landscape the exact
stationary solution of equations (\ref{ODE}) can be obtained by mapping them
into a polymer localization problem (Galluccio {\em et al} 1996). Recently,
a population genetics approach to the quasispecies model has been proposed
that, in spite of its simplicity, yields results that are qualitatively
similar to those obtained by solving the kinetic equations (Alves and
Fontanari 1996).

An alternative interpretation of the quasispecies model is given by
considering the $\kappa^\nu$ different strings $\vec{s}$ as different forms
(alleles) of a certain gene that determines the {\em fitness} $A_i$ of the
haploid organisms. Thus, this model is equivalent to the classical
one-locus, multiple-allele model of population genetics (Hartl and Clark
1989), except for the mutation mechanism which must be adapted to satisfy
the constraints imposed by the internal structure of the alleles.
Accordingly, Wiehe {\em et al} (1995) have generalized the original haploid
formulation of the quasispecies model so as to consider the evolution of
diploid organisms as well. An important by-product of that analysis is the
study of the effects of dominance on error thresholds, which has led to an
interesting conjecture about the evolution of dominance.

In this paper we employ the population genetics formulation of the
quasispecies model to investigate the error propagation in the reproduction
of diploid organisms. This approach allows us to study in great detail the
dynamical behavior of the model in the full space of the control parameters $%
\nu$ and $q$ as well as in the space of the parameters that specify the
fitness landscapes. We should mention that the analysis of Wiehe {\em et al}
(1995) was based on the numerical solution of the diploid counterpart of the
kinetic equations (\ref{ODE}) and on very crude approximations that neglect
the effects of back mutations.

In the population genetics formulation, the $\kappa^\nu$ different alleles
are grouped into $\left ( \nu + \kappa - 1 \right )!/\nu ! \left( \kappa - 1
\right )!$ classes, according to the number of monomers of each type they
have, regardless their specific position inside the allele. Hence, a given
class is characterized by the vector $\vec{P} = \left ( P_1,P_2, \ldots,
P_\kappa \right ) $, where $P_\alpha$ is the number of monomers of type $%
\alpha$ in any allele inside that class. Clearly, $\sum_\alpha^\kappa
P_\alpha = \nu$. The alleles belonging to the same class are assumed to be
equivalent, in the sense that their presence confers the same fitness value
on the genotypes. The crucial simplifying assumption of the population
genetics approach is that, given the monomer frequencies in generation $t$, $%
p_\alpha (t)$ with $\sum_\alpha^\kappa p_\alpha (t) = 1$, the frequencies of
alleles in class $\vec{P}$ are given by the multinomial distribution 
\begin{equation}  \label{multinomial}
\Pi_t (\vec{P}) = C_{\vec{P}}^\nu \left [ p_1(t) \right ]^{P_1} \left [
p_2(t) \right ]^{P_2} \ldots \left [ p_\kappa(t) \right ]^{P_\kappa}
\end{equation}
where $C_{\vec{P}}^\nu = \nu !/P_1! P_2! \ldots P_\kappa!$. Thus, at
generation $t$, the monomers are sampled with replacement from an urn
containing $\kappa$ different types of monomers in the proportions $p_\alpha
(t); \alpha = 1, \ldots, \kappa$.


Let $A(\vec{P}^i,\vec{P}^j) = A(\vec{P}^j,\vec{P}^i)$ denote the fitness of
the genotypes $\vec{P}^i \vec{P}^j$, i.e., genotypes composed of any pair of
alleles belonging to classes $\vec{P}^i$ and $\vec{P}^j$. Then the fraction
of monomers $\alpha$ that the genotype $\vec{P}^i \vec{P}^j$ contributes to
generation $t + 1$ is proportional to the product of three factors: (a) its
frequency in the population $\Pi_t (\vec{P}^i,\vec{P}^j)$, (b) its fitness $%
A(\vec{P}^i,\vec{P}^j)$, and (c) the average number of monomers $\alpha$
that replicate correctly, $q (P_\alpha^i + P_\alpha^j)$, plus the average
number of monomers $\beta \neq \alpha$ that mutate to $\alpha$, $%
[(1-q)/(\kappa-1)] \sum_{\beta \neq \alpha} (P_\beta^i + P_\beta^j)$. A
simple calculation yields the following equations for the time evolution of
the monomer frequencies: 
\begin{equation}  \label{sex}
p_\alpha (t+1) = \frac{1}{\kappa - 1} \left [ 1-q + \frac{\kappa q - 1}{2 \,
w_t} \sum_{\vec{P}^i}\sum_{\vec{P}^j} \Pi_t (\vec{P}^i,\vec{P}^j) A (\vec{P}%
^i,\vec{P}^j) ( P_\alpha^i + P_\alpha^j ) \right ] ,
\end{equation}
where 
\begin{equation}
\Pi_t (\vec{P}^i, \vec{P}^j) = \Pi_t (\vec{P}^i) \Pi_t (\vec{P}^j)
\end{equation}
and the normalization factor, 
\begin{equation}
w_t = \nu \sum_{\vec{P}^i}\sum_{\vec{P}^j} \Pi_t (\vec{P}^i,\vec{P}^j) \, A (%
\vec{P}^i,\vec{P}^j) ,
\end{equation}
is the average fitness of the entire population. Here the notation $\sum_{%
\vec{P}}$ stands for $\sum_{P_1=0}^\nu \ldots \sum_{P_\kappa=0}^\nu \delta
\left ( \nu, \sum_\alpha^\kappa P_\alpha \right )$, where $\delta (k,l)$ is
the Kronecker delta. It is interesting to note that equation (\ref{sex}) is
identical to the equation governing the evolution of sexually reproducing
haploid organisms (Alves and Fontanari 1996).

In the remainder of this paper we will consider binary strings only. In this
case there are two types of monomers ($\kappa = 2$), so that the alleles are
characterized by a single parameter, namely, the number of monomers of type
1 they have, $P_1 \equiv P$. The extension of our analysis to larger values
of $\kappa$ is straightforward. To proceed further we must specify the
fitness of the genotypes $P^i P^j$. According to Wiehe {\em et al} (1995) we
consider the following diploid analogue of the single-peaked fitness
landscape: 
\begin{equation}  \label{A}
A(P^i,P^j) = \left \{ 
\begin{array}{ll}
\left( 1 + a \right )^2 & \mbox{if $P^i=P^j = \nu$} \\ 
\left ( 1 + a \right )^{2h} & 
\mbox{if $P^i = \nu$ and
                        $P^j \neq \nu$} \\ 
1 & \mbox{if $P^i \neq \nu$ and
                        $P^j \neq \nu$} ,
\end{array}
\right.
\end{equation}
where $a > 0$ is the parameter measuring the selective advantage of the
so-called {\em master} allele $P = \nu$, and $- \infty < h < \infty $ is the
dominance parameter. The master allele is completely dominant for $h=1$ and
completely recessive for $h=0$. For $h = 1/2$ we find $%
A(P^i,P^j)=A(P^i)A(P^j)$ and so there is no dominance. In this case equation
(\ref{sex}) reduces to the equation that governs the evolution of asexually
reproducing haploid organisms (Alves and Fontanari 1996). Thus the intervals 
$h \in [0,1/2) $ and $h \in (1/2,1] $ delimitate the regions of recessivity
and dominance, respectively, of the master allele. There are other cases of
interest as well: $h > 1$ models the phenomenon of heterosis or hybrid vigor
(heterozygote advantage), while $h < 0 $ models the phenomenon that occurs
at the early stages of speciation when hybrids are less viable (heterozygote
disadvantage).

Inserting equation (\ref{A}) into the recurrence equation (\ref{sex}) yields
the following equation for the frequency of monomers of type 1 in generation 
$t$, $p_1 (t) \equiv p_t$, 
\begin{equation}  \label{p_single}
p_{t+1} = 1-q + \left ( 2 q - 1 \right ) \, \frac{ \Lambda_1 p_t^{2\nu} +
\Lambda_2 \left ( p_t + 1 \right ) p_t^\nu + p_t } { \Lambda_1 p_t^{2\nu} +
2 \Lambda_2 p_t^\nu + 1 } ,
\end{equation}
where 
\begin{equation}
\Lambda_1 = \left ( 1 + a \right )^2 - 2 \left ( 1 + a \right )^{2h} + 1
\end{equation}
and 
\begin{equation}
\Lambda_2 = \left ( 1 + a \right )^{2h} - 1 .
\end{equation}

In figure 1 we present the steady-state frequencies of alleles, obtained by
solving the recursion equation (\ref{p_single}) with $p_0 \approx 1$, as a
function of the error rate per monomer $1 - q$ for $\nu = 10$, $a=2$, and
different values of the dominance parameter. In the case of perfect
replication accuracy ($1-q=0$), the fixed point $p^* = 0$ is always
unstable, while $p^* = 1$ is stable for $h \leq 1$ only. For $h > 1$, a
third (stable) fixed point $1/2 < p^* \approx 1$ appears signalling the
emergence of heterosis. For $h \leq h_c \approx 1.75$ there are two distinct
regimes: the {\em quasispecies} regime characterized by a population
dominated by the master allele and its close neighbors, and the {\em uniform}
regime where the $2^\nu$ alleles appear in the same proportion (clearly the
class $P = \nu/2$ is the most favored in this case). The error rate at which
the discontinuous transition between these two regimes takes place is termed 
{\em error threshold} $1-q_t$. As $h$ increases, the size of the jump at the
transition decreases till it disappears at a critical value $h = h_c$.
Beyond that value it is no longer possible to distinguish the two regimes.

To better characterize the error threshold transition we concentrate our
analysis on the nature of the fixed points $p_{t+1} = p_{t} = p^*$ which are
given by the real roots of $f(p) = 0$, where 
\begin{equation}  \label{fix}
f \left ( p \right ) = \Lambda_1 \left ( p - q \right ) p^{2 \nu} +
\Lambda_2 \left ( 3p - 2 p q - 1 \right ) p^\nu - \left ( 1 - q \right )
\left ( 1 - 2p \right ) .
\end{equation}
For small error rates this equation has only one real root which corresponds
to the stable fixed point $p^* \approx 1$ associated to the quasispecies
regime. As the error rate increases, a double root appears originating two
new fixed points: a stable one, $p^* \approx 1/2$, associated to the uniform
regime, and an unstable one that delimitates the basins of attraction of the
stable fixed points. These fixed points co-exist till the error rate reaches
the threshold value $1 - q_t$, where the stable quasispecies fixed point and
the unstable one coalesce. For larger error rates, equation (\ref{fix}) has
only one real root which corresponds to the uniform fixed point. Thus, the
error threshold transition can be easily determined by solving $f(p) = d
f(p)/dp = 0$ simultaneously for $p$ and $q = q_t$. As mentioned above, since
these equations have two solutions we must choose the one with the larger
value of $p$. The critical point $h_c$ is determined by tuning the value of $%
h$ so that the three real roots of (\ref{fix}) coincide, i.e., we have to
solve the three equations $f(p) = d f(p)/dp = d^2 f(p)/dp^2 = 0$
simultaneously for $p$, $q=q_c$ and $h=h_c$.

Using the prescriptions given above, we present in figure 2 the error
threshold transition lines as a function of $h$ for $\nu = 10$ and several
values of $a$. The error threshold $1-q_t$ is practically insensitive to
variations of $h$ for negative and small positive values of this parameter.
It reaches its minimal value around $h = 0.5$ (non-dominance regime) and
then increases quickly as the system enters the dominance region, $h > 0.5$.
We note the reentrant behavior of these transition lines: for certain values
of $q$, the system undergoes two discontinuous transitions as $h$ is
increased. The transition lines end at critical points, which are shown in
figure 3 for different values of $\nu$. It is interesting to note that only
for $\nu < 7$ (or more exactly $\nu < 6.93$) the critical lines touch the
axis $1 - q_c = 0$. So, in these cases, there are values of the dominance
parameter $h > 0 $ for which the error threshold transition never occurs.

Heretofore we have concentrated on the location of the error threshold as a
function of the dominance parameter. We turn now to the analysis of the
composition of the population at the steady state. It can be characterized
by the average normalized Hamming distance from the master allele which,
within the population genetics framework, is given simply by $1 - p^*$. This
quantity is shown in figure 4 as a function of the error rate for $\nu = 10$%
, $a = 0.5$ and several values of $h$. What is remarkable about this figure
is that there exists a value of the error rate $1 - q =1 - q_r \approx 0.017$
such that the fixed point $p^* \approx 0.912$ is independent of $h$. This
fixed point, however, becomes unstable for $h < h_s \approx 0.244$. Thus,
although recessivity leads to a higher concentration of the master allele
for $1-q < 1-q_r$, this allele is quickly lost from the population for
larger error rates. The main effect of dominance is to postpone the error
catastrophe at the price of reducing the concentration of the master allele
in the population. We note that at the inflection point $q = q_r$ the
effects of dominance and recessivity are reversed. This point can be easily
determined by setting to zero the coefficient of the term $(1 + a)^{2h}$ in
equation (\ref{fix}), namely, 
\begin{equation}  \label{rev}
g(p) = -2 \left ( p - q \right ) p^\nu + \left ( 3 - 2 q \right ) p - 1 ,
\end{equation}
and solving $g(p) = 0$ together with $f(p) = 0$ for $p$ and $q=q_r$. 

Both analyses, the location of the error threshold and the composition of
the population, indicate that dominance allows the master allele to resist
to higher replication error rates than in the case of non-dominance.
Actually, for sufficiently large $h$, it can even avoid the error
catastrophe. This finding has been proposed as a possible explanation for
the fact that the wild type, i.e., the allele that predominates in a
population and that is particularly well suited to its environment, is often
dominant: the dominant alleles might be the prevailing wild-type ones simply
because they can tolerate higher error rates (Wiehe {\em et al} 1995).

In summary, we have employed the population genetics approach to the
quasispecies model to investigate the error threshold catastrophe in the
evolution of diploid organisms. In order to enhance the non-trivial effects
of the imperfect replication accuracy of the organisms on the population
composition, we have focused on a simple diploid analogue of the
single-peaked fitness landscape. Two distinct steady-state regimes are
observed: the quasispecies regime where the information about the
environment, modelled by the fitness landscape, is preserved in the
population composition, and the uniform regime, where this information is
irreversibly lost. In the space of the parameters $1-q$ and $h$, these
regimes are separated by discontinuous transitions lines that terminate at
critical points, beyond which they become indistinguishable. We have found
that dominance ($h > 0.5$) can postpone or even avoid the error catastrophe.
It is interesting, however, that a recessive allele ($h < 0.5$) can do
better than a non-dominant one ($h \approx 0.5$).

To conclude, we mention that our results are in qualitative agreement with
those of Wiehe {\em et al} (1995). Since the population genetics approach of
the quasispecies model incorporates only a few essential features of the
original chemical kinetics formulation, this agreement gives a strong
evidence for the robustness of the main conclusion drawn from the model,
namely, the existence of an error catastrophe that limits the size of
self-replicating organisms. Rather than just a caricature of the original
model, the population genetics model presented in this paper may be viewed
as a simpler, alternative model for investigating the evolution of
self-replicating organisms, which may greatly facilitate the analysis of
difficult problems such as the error propagation in finite populations and
the effects of cooperation or catalysis among the evolving organisms.

\bigskip

\section*{Acknowledgment}

This work was supported in part by Conselho Nacional de Desenvolvimento
Cient\'{\i}fico e Tecnol\'ogico (CNPq).


\newpage

\section*{References}

\parindent=0pt \parskip=10pt

Alves D and Fontanari J F 1996, {\em Phys. Rev. E} {\bf 54} 4048

Eigen M 1971, {\em Naturwissenchaften} {\bf 58} 465

Eigen M and Schuster P 1979, {\em The Hypercycle - A Principle of Natural
Self-Organization} (Springer-Verlag, Berlin)

Eigen M, McCaskill J and Schuster P 1989, {\em Adv. Chem. Phys.} {\bf 75} 149

Franz S, Peliti L and Sellitto M 1993, {\em J. Phys. A: Math. Gen.} {\bf 26}
L1195

Galluccio S, Graber R and Zhang Y-C 1996, {\em J. Phys. A: Math. Gen.} {\bf %
29} L249

Hartl D L and Clark A G 1989, {\em Principles of Population Genetics}
(Sinauer Associates, Sunderland)

Kauffman S A 1993, {\em The Origins of Order} (Oxford University Press,
Oxford)

Leuth\"{a}usser I 1986, {\em J. Chem. Phys.} {\bf 84} 1884

Leuth\"{a}usser I 1987, {\em J. Stat. Phys. }{\bf 48} 343

Tarazona P 1992, {\em Phys. Rev. A} {\bf 45} 6038

Wiehe T, Baake E and Schuster P 1995, {\em J. Theor. Biol.} {\bf 177} 1

\newpage

\section*{Figure captions}

\parindent= 0pt

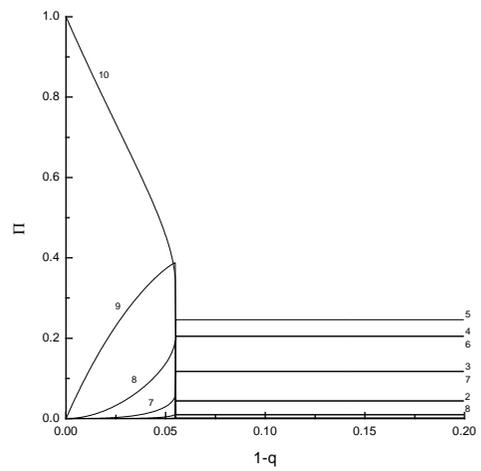
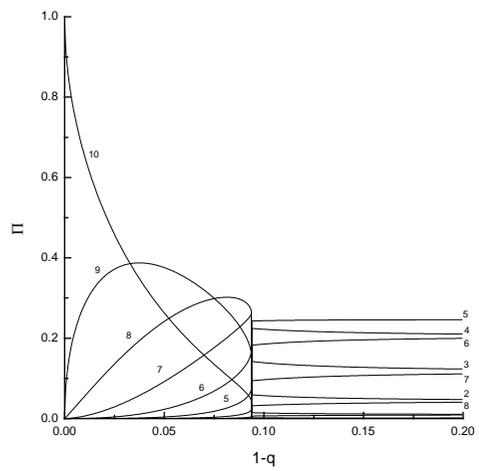
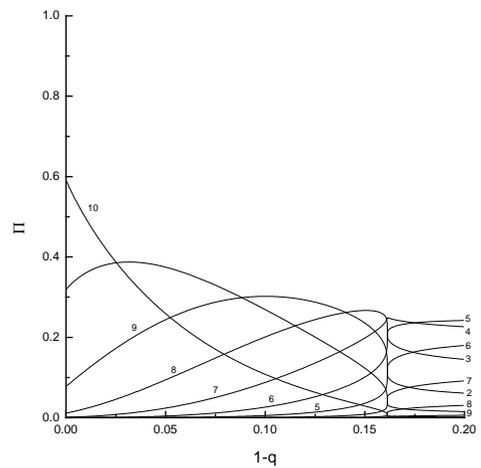
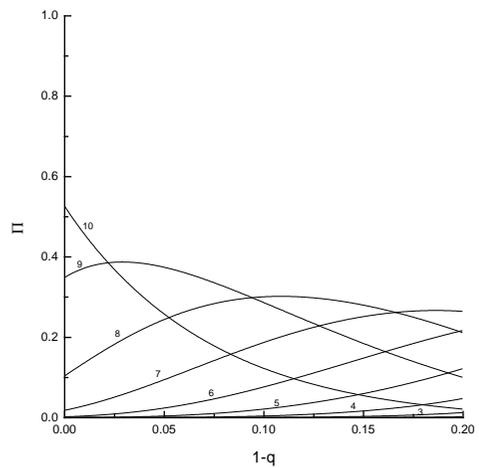
{\bf Fig.\ 1} Steady-state frequencies of alleles belonging to classes $P =
10$ (master allele) to $P = 0$ as a function of the error rate per digit $1
- q$ for $\nu=10$, $a=2$, and $(a)$ $h =0$, $(b)$ $h =1$, $(c)$ $h =1.5$,
and $(d)$ $h =2$.

\bigskip

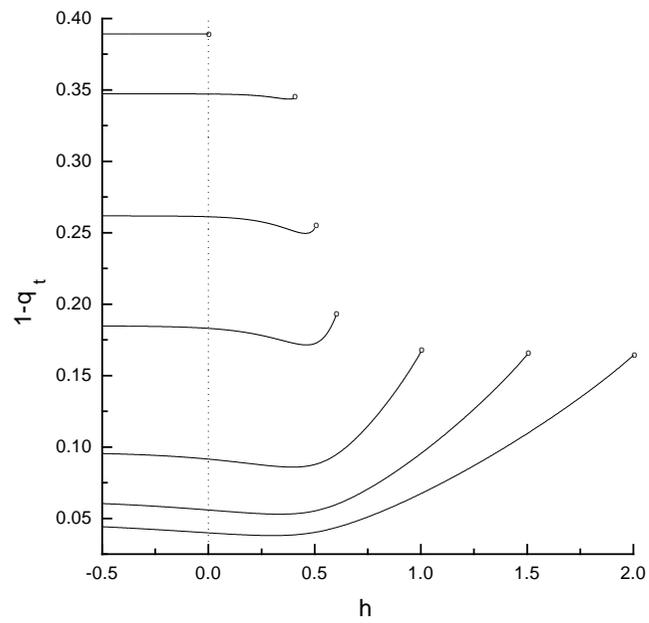
{\bf Fig.\ 2} Error threshold $1-q_t$ as a function of the dominance
parameter $h$ for $\nu = 10$ and (from top to bottom) $a = 314.8$, $186.1$, $%
57.0$, $18.8$, $4.4$, $2.0$, and $1.3$. The parameter $a$ was chosen so that
the transition lines end at critical points located at $h =0$, $0.4$, $0.5$, 
$0.6$, $1.0$, $1.5$ and $2.0$, respectively.

\bigskip

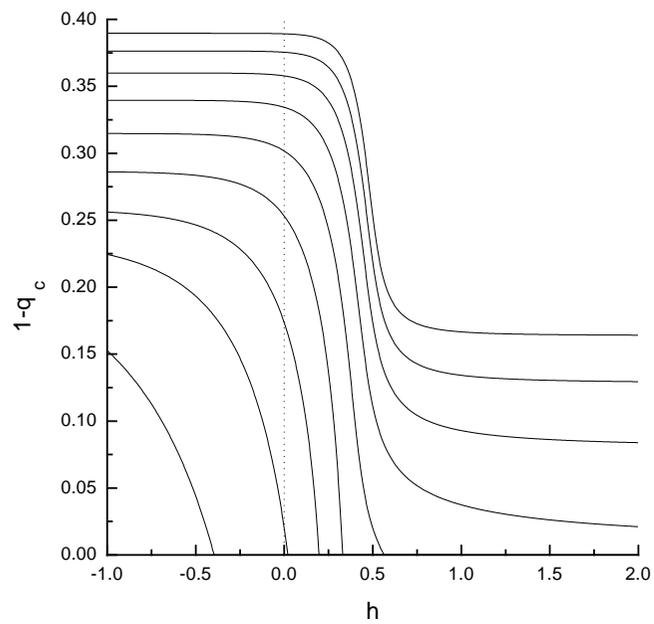
{\bf Fig.\ 3} Error threshold at the critical point $1-q_c$ as a function of
the dominance parameter $h$ for (from top to bottom) $\nu = 10$ to $\nu = 2$.

\bigskip

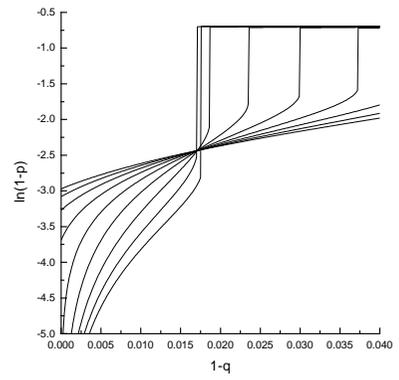
{\bf Fig.\ 4} Average normalized Hamming distance from master as a function
of the replication error rate for $\nu =10$, $a = 0.5$, and (from top to
bottom before the intersection) $h = 2$, $1.75$, $1.5$, $1.25$, $1$, $0.75$, 
$0.5$, $0.25$, and $0$. The curves for $h > h_s = 0.244$ intersect at the
inflection point $1-q_r = 0.017$.

\end{document}